\documentclass[12pt,onecolumn,journal]{IEEEtran}

\usepackage{ifpdf}
\usepackage[utf8]{inputenc}
\usepackage{resizegather}

\usepackage{cite}

\usepackage{graphicx}
\ifCLASSINFOpdf

\else

\fi

\usepackage{amsfonts}
\usepackage{amsthm}

\usepackage{algorithm}
\usepackage{algorithmic}

\usepackage{mdwmath}
\usepackage{mdwtab}

\usepackage{eqparbox}

\usepackage{siunitx}
\usepackage{mathtools}
\usepackage[hyphens]{url}
\usepackage[hidelinks]{hyperref}
\hypersetup{breaklinks=true}
\usepackage{graphicx, caption}
\usepackage{subcaption}
\usepackage{breqn}
\usepackage{cuted}
\usepackage{xcolor,cite, etoolbox}

\usepackage{url}

\usepackage{tabularx}
\usepackage{booktabs}
\usepackage{multirow}
\usepackage[final]{changes}

\makeatletter
\pretocmd\@bibitem{\color{black}\csname keycolor#1\endcsname}{}{\fail}
\newcommand\citecolor[1]{\@namedef{keycolor#1}{\color{blue}}}
\makeatother

\theoremstyle{plain}

\author{Stefano~Bonafini,
        Claudio~Sacchi~\IEEEmembership{Senior Member,~IEEE,}
        Riccardo~Bassoli~\IEEEmembership{Senior Member,~IEEE,}
        Fabrizio~Granelli~\IEEEmembership{Senior Member,~IEEE,}
        Koteswararao Kondepu~\IEEEmembership{Senior Member,~IEEE,}
        and Frank~H.P.~Fitzek ~\IEEEmembership{Fellow,~IEEE}

\thanks{S. Bonafini, C. Sacchi and F. Granelli are with the Department of Information Engineering and Computer Science (DISI), University of Trento, Trento, Italy.
(e-mail: \{stefano.bonafini,claudio.sacchi,fabrizio.granelli\}@unitn.it).}%
\thanks{K. Kondepu is with the Department of Computer Science and Engineering, Indian Institute of Technology Dharwad, Dharwad, India.
(e-mail:k.kondepu@iitdh.ac.in)}
\thanks{R. Bassoli and F. H.P. Fitzek are with the Deutsche Telekom Chair of Communication Networks, Institute of Communication Technology, Faculty of Electrical and Computer Engineering, Technische Universität Dresden, Dresden, Germany.
\par F. H.P. Fitzek is also with Centre for Tactile Internet with Human-in-the-Loop (CeTI), Cluster of Excellence, Dresden, Germany.
(e-mail: \{riccardo.bassoli,frank.fitzek\}@tu-dresden.de).}%
}

\begin{document}

\title{\added{An Analytical Study on Functional Split in Martian 3D Networks}


}
\maketitle

\begin{abstract} 
As space agencies are planning manned missions to reach Mars, researchers need to pave the way for supporting astronauts during their sojourn. This will also be achieved by providing broadband and low-latency connectivity through wireless network infrastructures. 
In such a framework, we propose a Martian deployment of a 3-Dimensional (3D) network acting as Cloud Radio Access Network (C-RAN). The scenario consists, mostly, of unmanned aerial vehicles (UAVs) and nanosatellites. Thanks to the thin Martian atmosphere, CubeSats can stably orbit at very-low-altitude. This allows to meet strict delay requirements to split functions of the baseband processing between drones and CubeSats. The detailed analytical study, presented in this paper, confirmed the viability of the proposed 3D architecture, under some constraints and trade-off concerning the involved Space communication infrastructures, that are discussed in detail.

\end{abstract}


\section{Introduction}
The new horizon of space exploration is to land a human crew on Mars around 2026, by accepting a certain degree of risks~\cite{HumanonMars}. Some other forecasts state that, in more or less a hundred years, we shall be able to set a base on the Martian surface to host up to 50 persons~\cite{BaseonMars}. In such a perspective, researchers need to tackle multifaceted problems before being able to move people on Mars: from the design and development of the long-journey vehicle to the support of the astronauts during their sojourn. To this aim, it is of paramount importance to support human life and operations on Mars through the deployment of efficient connectivity infrastructures. 
This will allow the exchange of real-time information or emergency messages, post-processing of data, navigation and remote control of rovers, landers and UAVs. Experiments conducted by NASA demonstrated that delays and limited bandwidth in extra-terrestrial missions would produce frustration and uncertainty in the crew~\cite{SocialityonMars2}. On the other hand, the presence of efficient connectivity would improve sociality, the quality of life and, as consequence, increase the probability of success of the mission. 

In the recent literature, few works have been focused on the provision of efficient mobile connectivity on Mars. Some contributions are mainly devoted to observe the wireless propagation on Mars in the context of machine-to-machine communications~\cite{1367731} and~\cite{1559461}. Other papers preliminarily analyzed the functional viability of porting terrestrial LTE (and LTE-A) services on Mars obtained by installing eNodeBs somewhere on the planet surface~\cite{LTEonMars}~\cite{LTEonMars2}. Despite the interesting and useful results presented in these works, such approaches do not still provide reliable technical answers to the connection requirements of a future manned mission.

Thus, in this piece of research, we study the deployability of "Beyond 5G" wireless infrastructure in the framework of Martian communications, where heterogeneous devices, such as UAVs and nanosatellites, constitute the C-RAN and huger platforms in higher orbits become the core network. Connectivity is, hence, offered from above through a 3D network - one of the driving pillars for the upcoming terrestrial 6G~\cite{3DNet6G} - on Mars. The advantages taken by the availability of such a broadband degree of connectivity are supported by various considerations: low end-to-end (E2E) latency, huge bandwidth availability and high energy efficiency, which is attested around $90\%$ more than the 4G one~\cite{5GBenefits2}. With respect to an on-ground fixed architecture, the advantages are increased coverage, on-demand anywhere-anytime connectivity, high degree of reconfigurability and lower outage probability, which on the contrary would be strongly experimented due to the heavily rocky and craterized nature of Mars~\cite{9438180}. Moreover, in a futuristic perspective, by supposing to have many UAVs and CubeSats, respectively, flying and orbiting on Mars, we should be able to integrate them in the same space ecosystem and coordinate them with on-ground devices. In such a perspective, a space habitat could be interconnected with machines far away, while the human personnel inside could easily acquire data from them.

The problem to be solved is how to virtualize the necessary network functionalities into flying and orbiting non-dedicated hardware present in such a hostile environment, which will be conveniently equipped with transceivers, antennas, processing units and other scientific payload.
The way toward a viable solution to this problem may be represented by the work of Bassoli \emph{et al.}~\cite{Bassoli2020}, where the feasibility of advanced functional splitting and C-RAN solutions has been studied in a challenging terrestrial scenario, related to secure and efficient border monitoring. The authors of~\cite{Bassoli2020} and, especially~\cite{9172316}, deeply analyzed and also simulated the so-called ``Split D" between UAVs (hosting the LTE radio headend) and CubeSats. The results open up to the possibility of customizing and upgrading such an architecture, by meeting stricter requirements imposed by the 5G standard, to assure connectivity, whenever and wherever is needed. For the aforesaid motivations, we believe that 3D network solutions may deserve study and consideration also in the Martian context.

\begin{figure*}[t]
\centering
\includegraphics[width=1.0\columnwidth]{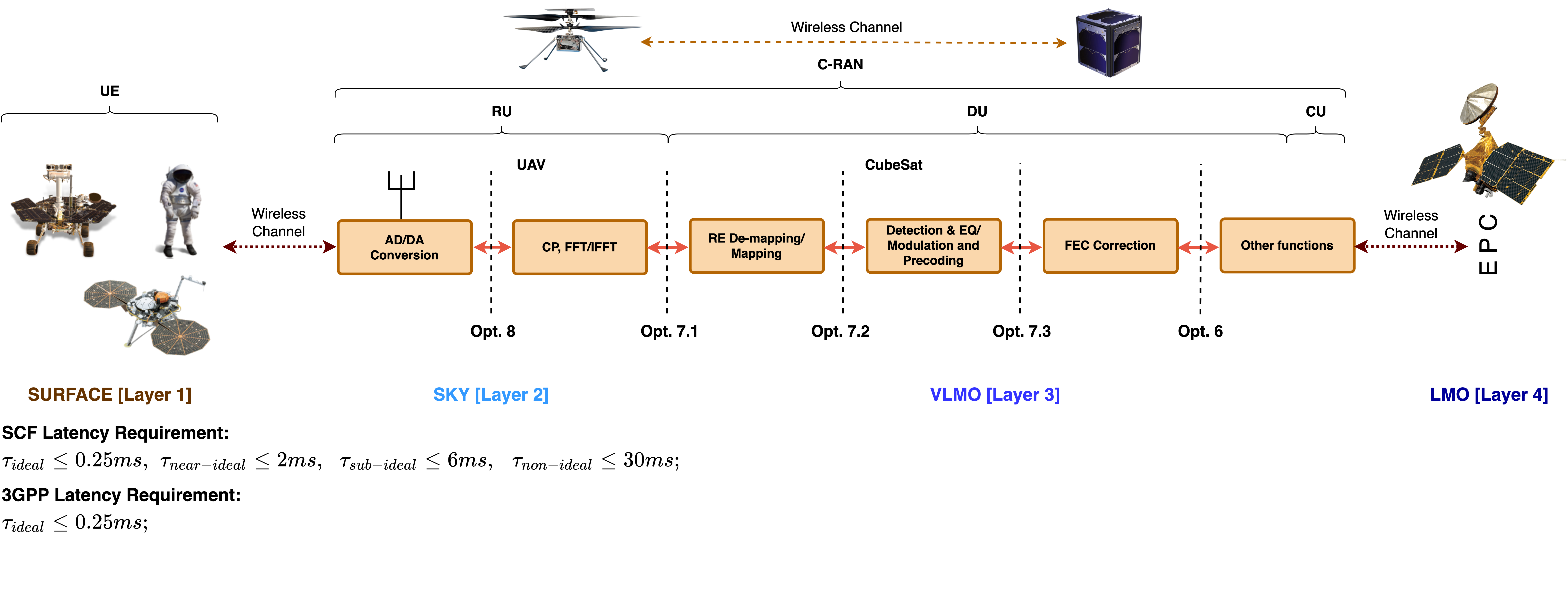}
\caption{Picture representing the overall network and the splitting options under study.}
\label{split}
\end{figure*}

The present work will investigate in an analytical framework the main feasibility constraints related to a Martian C-RAN architecture for B5G 3D network, based on UAVs and CubeSats. In detail, the splitting options characterized by the most stringent specification in terms of latency (option 7.3-8 shown in Fig.~\ref{split}) will be analyzed by computing the required CubeSat altitude with respect to the propulsion force needed to mantain the orbit. The next step will be the estimation of the elevation angle and the resulting session time available from CubeSat and the hovering UAV, while fixing the altitude. Consequently, we shall correlate drag force and session time to find the best CubeSat orbit altitude and provide the main architectural parameters. To conclude the analysis, few considerations will be raised for supposing the deployment of CubeSats into higher orbits to sensibly increase their lifetime, thus assuring service continuity over time even without continuous orbit corrections.  

To the best of our knowledge, no other contribution published in the open literature presented such kind of analysis, applied to a non-terrestrial context.

The rest of the correspondence is structured as follows: sect.II will describe the Martian 5G scenario, sect.III will detail the proposed methodology, sect.IV will present selected simulation results,  sect.V will draw paper conclusions.

\section{The Scenario}
\subsection{Overall architecture}
As written above, we believe in the necessity of providing robust, broadband and global connectivity to serve computing machines, humans and any other kind of sensors, i.e: the \emph{user equipment} (UEs) of the network, over the Martian surface. For this reason, we propose an ad-hoc network composed by four layers: a surface layer, made of the UEs, and an aerial layer, which will mainly be devoted to take in charge the remote radio unit (RU) functions of the base station (BS). Next, a very-low Mars orbit (VLMO) layer, in which constellations of CubeSats will take the responsibility of running distributed units (DUs) and centralized units (CUs) functions, and, concluding, a low Mars orbit (LMO) layer, with larger orbiters as the Mars Reconaissance Orbiter (MRO), hosting the evolved packet core (EPC) of the B5G network and providing backhaul links with aerostationary orbiters, in their turn, communicating with Earth to forward and receive data. In this work, we concentrate our focus on the network setup concerning the aforesaid nodes and, especially, in the challenging UAV-to-CubeSat Martian link where the baseband processing is split. The analysis of the issues inherent to the integration of the deployed local infrastructure with the long-haul interplanetary connection will be matter for future work.

The network layers are vertically and horizontally interconnected, meaning that a communication happens sending data through the uplink path from UE to UAV, i.e. the RU, from UAV to CubeSat, i.e. the DU and CU, and from CubeSat to a higher orbiter, i.e. the EPC. On the other hand, UEs can download information, which are delivered by the EPC following the reverse route from orbit to surface. Moreover, the nodes within a layer could exchange data between themselves, for instance, to counteract against possible failures.

Service continuity is enabled through the handover between serving satellites, i.e. CubeSats covering UAVs and orbiters covering CubeSats. Among the various techniques detailed in the state-of-the-art, the slow handover mode described in~\cite{9566290} well cope with the requirements of the proposed application. A UAV associates to the satellite closest to the Zenith for the best conditions in terms of delay and pathloss. Once the satellite comes to the position of \textit{loss-of-signal}, occurring when delay requirements are no longer met, the user re-links with the next satellite approaching the Zenith. The same happens for LMO orbiters, which are deployed in low circular orbits to reduce the contact distances between them and CubeSats. Although the fast handover mode could overall guarantee a better link budget~\cite{9566290}, the slow handover mode offers, for the same number of satellites, a higher session duration $t_{s}$ that is advisable for our scenario. We will see in the following a detailed evaluation of such a parameter.

Besides the couple of antennas and the processing unit mounted to operate network functions and communicate with the on-ground UEs and with CubeSat constellation, UAVs will be equipped with solar panels recharging lithium-ion batteries. However, due to their reduced size and weight and the thin Martian atmosphere ($\frac{1}{100}$ of the terrestrial atmospheric density~\cite{AtmDensity1}), most of the available power has to be used to make the UAV flying for a considerable range of time. 
Therefore, battery life is a crucial parameter to be optimized, and so, the processing unit should consume as less as possible, while operating for maintaining a continuous and reliable connection with the UEs. It should be remarked here that experiments related to Martian drones are in a very early stage and, so far, have been only devoted at demonstrating the possibility for small remotely-controlled rotorcrafts to fly in the thin Mars atmosphere \cite{Ingenuity}. Such experiments will achieve further goals in the near future that may enable future Martian drones to embark light but vital payloads, i.e.: to survive the cruise to Mars, to autonomously charging themselves with their solar panels and to communicate to and from the helicopter via a subsystem called Mars helicopter base station \cite{Ingenuity}.

\begin{figure}[t]
\centering
\includegraphics[width=0.75\columnwidth]{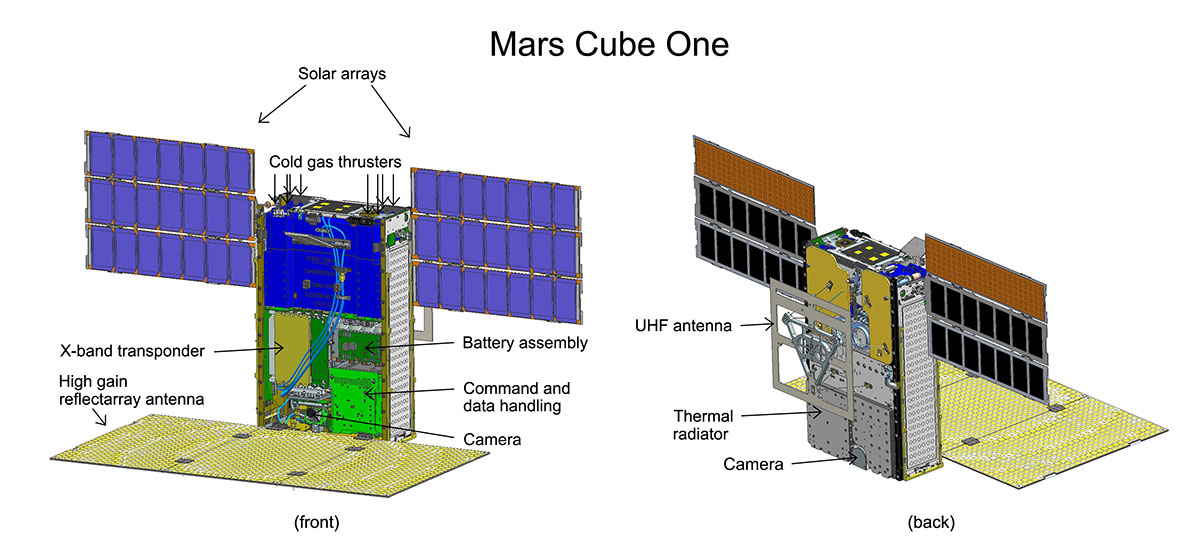}
\caption{MarCO CubeSats (courtesy by \cite{MarCOdemo}, source: JPL).}
\label{marco}
\end{figure}

However, virtualizing BBU (v-BBU) functions will remain a heavy processing task, even for advanced and efficient UAV systems. For this reason, we aim to move the most complex functions of the v-BBU on CubeSats (options 7.3-8~\cite{8479363} shown in Fig.~\ref{split}). Clearly, these CubeSats should be characterized by size greater than the typical cubic-shaped unit (1U) of $10x10x10cm$ in order to embark the scientific payload along with the orbit maintenance system. The remaining available room on the CubeSat will be filled with computing and storage units. The energy resources should be harvested thanks to solar arrays and lithium-ion batteries. The recent development of \emph{MarsCubeOne (MarCO)} mission, targeted at sending swarms of CubeSats on Mars, is indeed designing small satellites supplied by solar panels. Their size and weight look suitable to host advanced payloads. Indeed, MarCO's design is a six-unit CubeSat. Each of the two platforms has a stowed size of 36.6 centimeters by 24.3 centimeters and by 11.8 centimeters \cite{MarCOdemo} (see Fig.\ref{marco}). 
So far, the results of the MarCO mission have been contrasting. MarCOs satellites, named EVE and WALL-E, served as communication relays during the \emph{InSight} rover landing, beaming back data at each stage of its descent to the Martian surface in near-real-time \cite{MarCO2}. WALL-E sent some remarkable images of Mars, while EVE performed some radio science experiments. The last contact with the MarCO pair was in early 2019. The NASA mission team investigated the reasons for why they haven't been able to contact the pair. WALL-E should have experienced problems due to a leaky thruster along with some control issues. Moreover, the brightness sensors that allow the CubeSats to stay pointed at the Sun and recharge their batteries could be another failure point. However, as claimed in \cite{MarCO2}, the mission was always about pushing the limits of miniaturized technology and seeing just how far it could take. In early launches, MarCO satellites demonstrated to be capable of orbiting and transmitting/receiving signals. Future versions are expected to go farther in advanced radio system experimentation, of course after solving the problems inherent to the platform and the control system that hindered the early phase of the mission. In such a perspective, the possibility of embarking a processing payload capable of supporting more advanced tasks, like v-BBU ones, might become more actual. 

\subsection{Splitting options}
The interest around the various splitting options is due to the possibility of enabling a complete virtualization of BBU tasks in non-dedicated hardware, as it could be a reconfigurable processing unit installed on a CubeSat or UAV.

The Third Generation Partnership Project (3GPP) identified eight possible functional split options, as shown in Fig.~\ref{split}. Here, split option 1 (i.e., opt.1) detaches the Packet Data Convergence Protocol (PDCP) from the network layer of Radio Resource Control (RRC), while opt.2 does the same between the Radio Link Control (RLC) and the PDCP. Opt.3 is operated within the RLC and opt.4 divides the Media Access Control (MAC) from the RLC, while Opt.5 separates the Lower MAC from the Upper MAC. The last splitting operation within the datalink layer is opt.6, which is done before the Forward Error Correction (FEC). From opt.7-8 we move to the physical (PHY) layer. Opt.7.3 is performed between the detection, equalization, modulation, precoding and the FEC, while opt.7.2 goes deeper into the PHY-layer detaching the resource element (RE) mapping, or demapping, functions. The CP, insertion or removal, and Fast Fourier Transform (FFT), or Inverse FFT (I-FFT), is implemented in RU if opt.7.1 is considered. At the end, opt.8 is considered just after the analog-to-digital (AD), or digital-to-analog (DA), conversion. Note that bandwidth and latency requirements should be met when considering these splits as specified in 3GPP TR 38.801~\cite{3GPP}.
The achievement of the fronthaul data-rate needed to guarantee the fulfilment of the splitting functionalities depends on many factors, such as the number of antennas at RU side, the available sampling rate of the analog front-end and the selected split option. A detailed study regarding such aspects is carried out in \cite{Bonafini2021Aerospace}, where the fronthaul communication system coping with the requirements of different v-BBU splitting options has been thoroughly described and technical solutions for its practical implementation have been proposed. The latency requirements are variously imposed by the standardization bodies upon specific criteria. For instance, the Small Cell Forum (SCF) defines the one-way maximum latency $\tau_{ideal}=0.25ms$ as for the ideal case,  $\tau_{near-ideal}=2ms$ for the near-ideal case, $\tau_{sub-ideal}=6ms$ and $\tau_{non-ideal}=30ms$, respectively, for the sub-ideal and non-ideal case~\cite{SCF}. For the 3GPP \cite{3GPP}, the latency for splitting opt.6-8 should be less than $0.25ms$, while for opt.5 should be of the order of hundreds of microseconds, for opt.4 around $0.1ms$, for opt.2-3 ranging from $1.5ms$ to $10ms$ and, finally, for opt.1, equal to $10ms$. The authors of~\cite{8479363} state that, for opt. from 1 to 5, the requirements to be accounted for designing the architecture and the communication links are way relaxed with respect to the ones needed for opt. from 6 to 8. It seems, then, that the critical latency is $\tau_{ideal}=0.25ms$~\cite{Chaudhary}. This last will be assumed as our reference value. 
Allowing splitting opt.8, which however could be unfeasible due to the prohibitive fronthaul data rate, is meant to ensure the possibility of meeting most of the requirements of the other splitting operations. For this reason, we find more interesting to deal with opt.7-8, at least for what concerns our application. As said before, our aim is to consider the RU supported by the UAVs, while most of the DU and the CU functions are implemented on the CubeSat.

Power consumption aspects may result critical in the fulfilment of such kind of advanced networking tasks. In this paper, we shall provide some notes about the concerned issues, leaving for future work further detailed analysis. Citing the literature, a dual-site virtualized RAN (vRAN) model has been described in~\cite{Ismail}, where remote and central sites are utilized to split the RAN functionalities. The RAN components at the central site are assumed to be virtualized. The authors of~\cite{Ismail} reported detailed bandwidth, power consumption and energy efficiency metrics as a function of functional split percentage deployed at the respective sites.
Furthermore, the virtualization of RAN functions is expected to provide more advantages, such as energy consumption due to dynamic resource allocation and load balancing as described in~\cite{ETSI}. In~\cite{Justine}, the authors reported that a significant energy savings could be obtained by offloading RAN functions such as CP, FFT, and I-FFT to a field programmable gate array (FPGA). Note that changing the functional split option from opt.7.1 to opt.7.2x at the RU, provides a variable bit rate by further reducing fronthaul bitrate.


In the following of the paper, the discussion will go to the direction of ensuring the needed ideal latency $\tau_{ideal}$ of about $250\mu s$, which will rise few interesting considerations on the possibility of lowering CubeSat's orbits due to the Martian intrinsic physics and atmosphere, in order to allow low-latency applications for future missions. 

\section{Proposed methodology}
The feasibility of the splitting options 7-8 is assessed by meeting the latency requirement in terms of $\tau_{ideal}=250\mu s$~\cite{3GPP}. Thus, to support low-latency applications in the order of 1-10 $ms$, $\tau_{ideal}$ should be our reference in the system design. This means that the distance $d$, lately referred as slant range, between UAV and CubeSat (CS), in a point-to-point communication, can range up to $d_{max}\sim75km$, given that $d=c \cdot \tau_{ideal}$ with $c$ the speed of light. 
By starting from this assumption, the proposed methodology can be resumed in the points listed below:
\begin{enumerate}
    \item\label{first} evaluating the Martian atmospheric density $\rho$ by sweeping the altitude value; 
    \item\label{second} computing the approximate drag force $F_{drag}$ over 1U, 6U, 12U CubeSat, thus understanding the needed propulsion force $F_{prop}$ to correct and maintain the orbit;
    \item\label{third} analyzing the acceptable elevation angle $\epsilon$, while computing the slant range $d$ between the hovering UAV and the orbiting CubeSat, to assure $\tau_{ideal}$;
    \item\label{fourth} obtaining the maximum session time $t_{s}$ between UAV and CubeSat.
\end{enumerate}
First of all, we estimate the Martian atmospheric density $\rho(h_{CS})$ through the model in~\cite{2018EPSC}, i.e.:
\begin{equation}
    \rho(h_{CS}) = \rho_0\cdot e^{\frac{-h_{CS}}{H}}
    \label{eq1}
\end{equation}
where $H=11.1km$ the atmosphere's scale height~\cite{AtmDensity2}, $h_{CS}$ the CubeSat actual altitude and $\rho_0$ two reference densities, a low one $\rho_0=0.0001\frac{kg}{m^3}$ and a high one $\rho_0=0.001\frac{kg}{m^3}$.
The drag force $F_{drag}$ is expressed as below~\cite{2018EPSC}:
\begin{equation}
    F_{drag}(h_{CS})=\frac{1}{2}\bigg(\rho(h_{CS})\cdot v_{CS}(h_{CS})^2\cdot C_{D}\cdot A_{CS}\bigg)
    \label{eq2}
\end{equation}
and computed accordingly to different size of CubeSat. Indeed, $v_{CS}$ stands for the circular velocity of CubeSat, which is the following~\cite{lissauer2013fundamental}:
\begin{equation}
    v_{CS}(h_{CS}) = \sqrt{\frac{G\cdot M_{Mars}}{h_{CS}+R_{Mars}}}
    \label{eq3}
\end{equation}
with $G=6.67\times10^{-11}\frac{Nm^2}{kg^2}$ the gravitational constant, $M_{Mars}=6.39\times10^{23}kg$ and $R_{Mars}=3389.5\times10^{3}m$, respectively, the planet mass and radius~\cite{AtmDensity2}, 
$C_{D}=2.0$ is the drag coefficient, $A_{CS}$ is the CubeSat cross-section obtained by the formulation, valid for parallelepiped-shaped spacecraft, in~\cite{CSA}:
\begin{equation}
    A_{CS} = \frac{1}{2}\bigg(S_{1}+S_{2}+S_{3}\bigg)
    \label{eq4}
\end{equation}
with $S_{1},\,S_{2},\,S_{3}$ the mean area of the visible CubeSat surfaces, which, however, does not consider the area occupied by the possible presence of the solar array. 
The considered dimensions are $10x10x10cm$ for 1U, $20x30x10cm$ for 6U and $20x30x20cm$ for 12U.

\begin{figure}
\centering
\includegraphics[width=0.7\columnwidth]{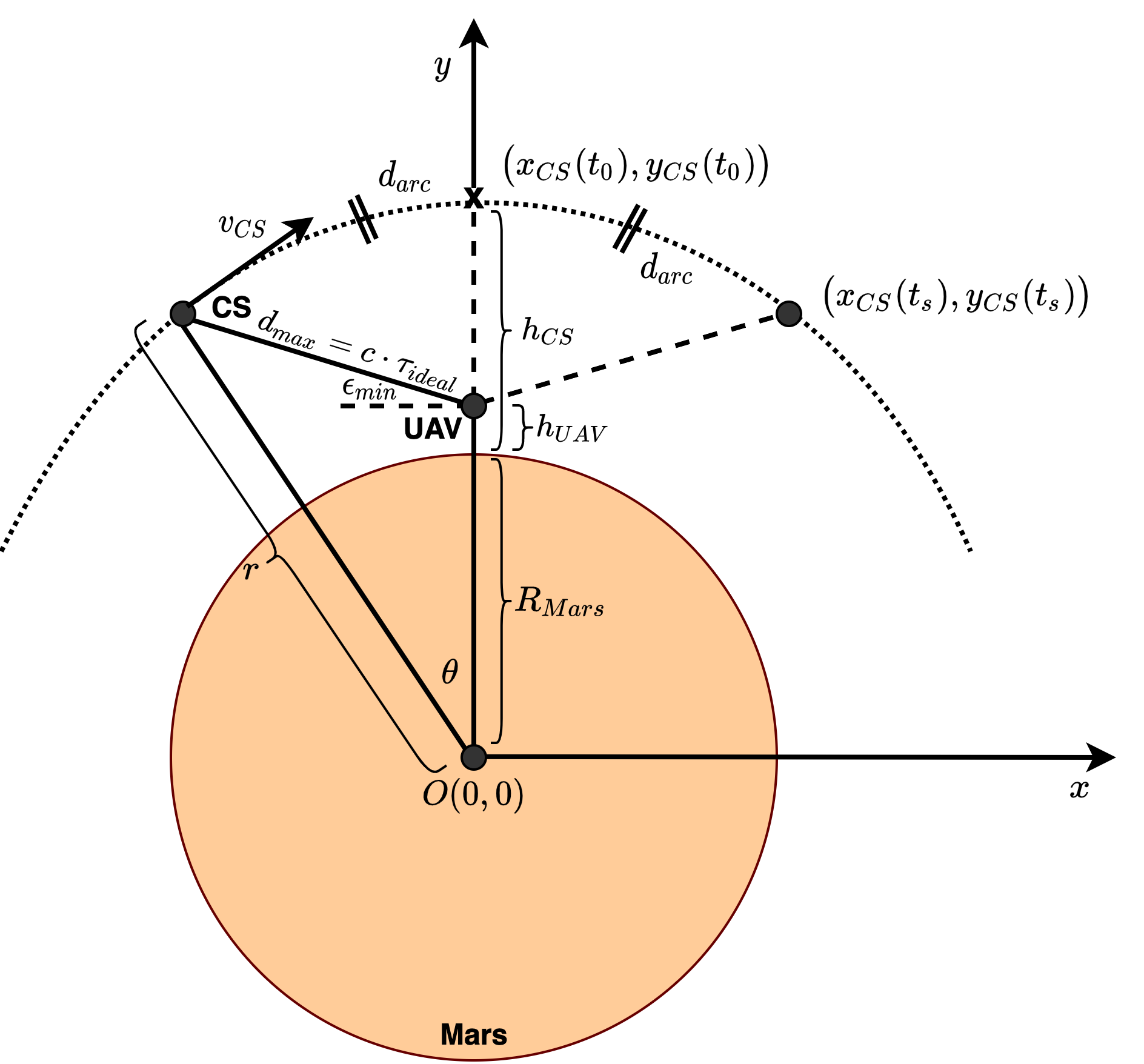}
\caption{\added{2D geometrical representation of the circular arc in which communication happens in session time $t_{s}$ between CubeSat and UAV.}} 
\label{geom}
\end{figure}

At this point in order to evaluate the minimum orbital altitude of CubeSat, under which it would be impossible to counteract the drag force and maintain the orbit, it is necessary to know the force $F_{prop}$, expressed by the thruster installed on the small satellite platform. $F_{prop}$ should be, at least, equal to $F_{drag}$ to be able to continuously correct the satellite orbit. 
Later, we will show some commercial and non-commercial thrusters and their impact on the minimum allowed altitude.
As briefly introduced in the list at the beginning of this section, the minimum altitude is the minimum acceptable distance $d_{min}$ between UAV and CubeSat. Indeed, while we consider for simplicity the UAV hovering on the Martian surface, 
CubeSat is circularly orbiting, in LMO or VLMO, around Mars at the velocity described in Eq.~\ref{eq3}. 
Consequently, CubeSat will be in the nearest point to the UAV only when the elevation angle, i.e. the angle between the line of sight (LOS) connecting CubeSat and UAV and the relative horizontal plane, is $\epsilon=\frac{\pi}{2}$. The LOS length $d$ is also called slant range and it is formulated by starting from the law of cosines, and customized to fit our problem, as follows~\cite{Cakaj2014}:
\begin{subequations}
\begin{gather}
\label{eq5}
\begin{multlined}
    (R_{Mars}+h_{CS})^2=\\(R_{Mars}+h_{UAV})^2+d^2-2(R_{Mars}+h_{UAV})cos(90+\epsilon)\end{multlined}\\
\begin{multlined}    
    (R_{Mars}+h_{CS})^2=\\(R_{Mars}+h_{UAV})^2+d^2+2(R_{Mars}+h_{UAV})sin(\epsilon)\end{multlined}\\
\begin{multlined}    
    \frac{(R_{Mars}+h_{CS})^2}{(R_{Mars}+h_{UAV})^2}=\\\bigg(1+\frac{d^2}{(R_{Mars}+h_{UAV})^2}+2\frac{sin(\epsilon)}{(R_{Mars}+h_{UAV})}\bigg)\end{multlined}\\
\begin{multlined}     
    \frac{(R_{Mars}+h_{CS})^2}{(R_{Mars}+h_{UAV})^2}=\\\bigg(cos^2(\epsilon)+sin^2(\epsilon)+\frac{d^2}{(R_{Mars}+h_{UAV})^2}+2\frac{sin(\epsilon)}{(R_{Mars}+h_{UAV})}\bigg)\end{multlined}\\  
\begin{multlined}       
    \frac{(R_{Mars}+h_{CS})^2}{(R_{Mars}+h_{UAV})^2}-cos^2(\epsilon)=\bigg(sin(\epsilon)+\frac{d}{(R_{Mars}+h_{UAV})}\bigg)^2\end{multlined}\\ 
\begin{multlined}     
    d=\left[\sqrt{\frac{(R_{Mars}+h_{CS})^2}{(R_{Mars}+h_{UAV})^2}-cos^2(\epsilon)}-sin(\epsilon)\right]\cdot(R_{Mars}+h_{UAV})\end{multlined}
\end{gather}
\end{subequations}
where $h_{UAV}$ is the drone height, as visible from Fig.\ref{geom}, and the elevation angle $\epsilon$ is defined in the range $[0,\,\frac{\pi}{2}]$. The minimum allowed elevation angle $\epsilon_{min}$ is found by searching, in the matrix representing the slant range $d$, the maximum LOS distance $d_{max}=c\cdot \tau_{ideal}$ between UAV and CubeSat for each $h_{CS}$ and $h_{UAV}$.
Now, thanks to trigonometric functions, the session time between UAV and CubeSat, i.e. the time that elapses between having $\epsilon=[\frac{\pi}{2},\,\pi-\epsilon_{min}]$, can be estimated. Such a range is inherent to the slow handover mode, as introduced in sect.II, where a UAV re-connects to a new CubeSat close to the Zenith when the distance from the previous CubeSat exceeds $d_{max}$. 
First, the angle $\theta$ in (rad) subtended by the circular minor arc, representing the orbit within $\epsilon=[\epsilon_{min},\,\frac{\pi}{2}]$ with radius $r=h_{CS}+R_{Mars}$ and center in the Earth's core, is found in the following manner:

\begin{subequations}
\begin{gather}
\begin{multlined}
    (R_{Mars}+h_{CS})sin(\theta)=d_{max}cos(\epsilon_{min})\end{multlined}\\
\begin{multlined}  
    \theta_{max} = asin\bigg(\frac{d_{max}\cdot cos(\epsilon_{min})}{R_{Mars}+h_{CS}}\bigg)\end{multlined}
    \label{eq6}
\end{gather}
\end{subequations}

Then, the session time $t_{s}$ in which UAV and CubeSat can communicate is given by the distance $d_{arc}$, i.e. the circular arc travelled by CubeSat divided by the orbital speed $v_{CS}$, i.e.:

\begin{subequations}
\begin{gather}
\begin{multlined}
    d_{arc}= \bigg(\theta \cdot (R_{Mars}+h_{CS})\bigg)\end{multlined}\\
\begin{multlined}    
    t_{s} = \frac{d_{arc}}{v_{CS}}\end{multlined}
    \label{eq7}
\end{gather}
\end{subequations}

To conclude and further clarify, $t_{s}$ is the session time achievable if we suppose to establish a communication between UAV at a certain altitude and CubeSat exactly moving from Zenith to the loss-of-signal position, where $d=d_{max}$. Thus, this estimate accepts a certain degree of approximation, which, however, seems reasonable for our purpose of assessing the feasibility of functional split in 3D networks.
\section{Results}
As previously mentioned in sect.III, by comparing the drag force $F_{drag}$ and the thrusters force $F_{prop}$, we are able to estimate the minimum allowed altitude for the 1U, 6U and 12U CubeSats.
\begin{table}[t]
\caption{CubeSat parameters.}
\label{tab:cs}
\centering
\resizebox{0.7\textwidth}{!}{  
\begin{tabular}{l|c|c|c|c}
\toprule
 & Form Factor & Size & Weight & Nominal Thrust \\ \hline 
\midrule
Commercial CubeSat & 1U & $10x10x10cm$ & $1.33kg$ & $1mN$ (off-the-shelf)\\ 
MarCO CubeSat & 6U & $20x30x10cm$ & $13.5kg$ & $(10x4)mN$ (Vacco's MiPS)\\ 
Hybrid CubeSat & 12U & $20x30x20cm$ & $25.0kg$ & $44.4N$ (JPL HT)\\
\bottomrule
\end{tabular}}
\end{table}
\begin{figure}[t]
\centering
\includegraphics[width=0.9\columnwidth]{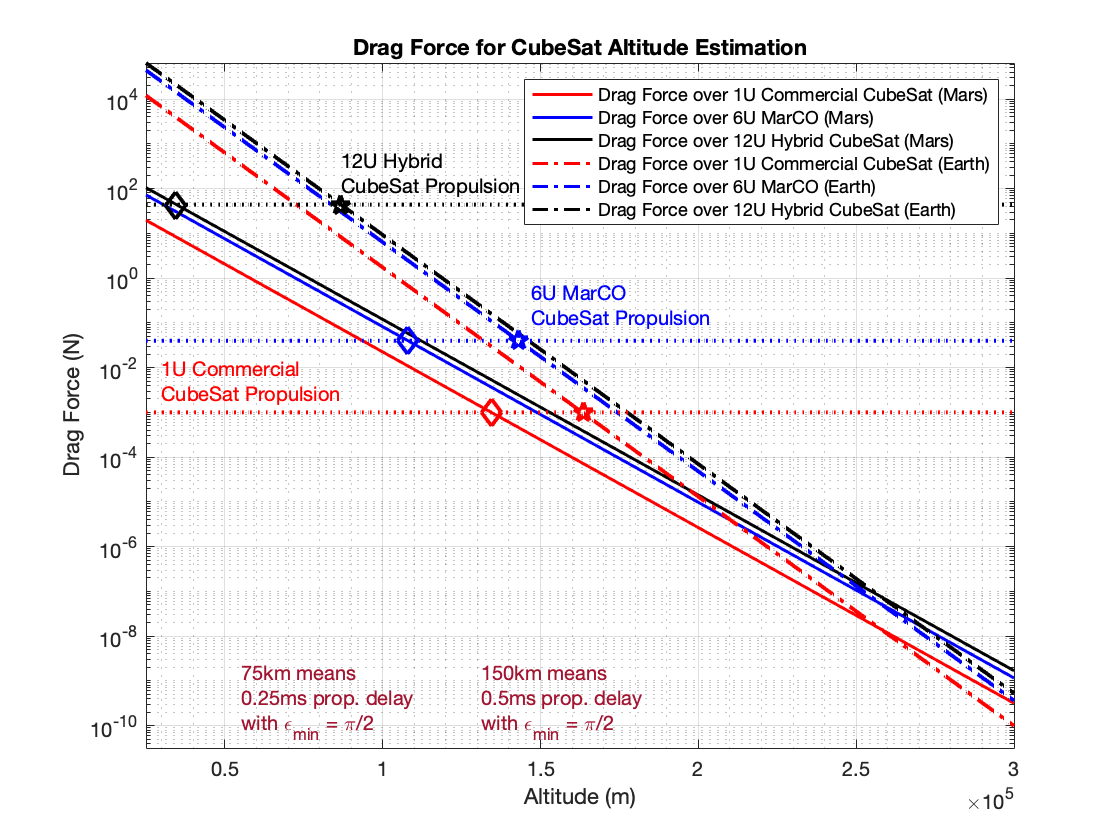}
\caption{Drag force $F_{drag}$ over 1U, 6U and 12U CubeSats on VLMO with respect the low Earth orbit (LEO). The dotted lines stands for the propulsion force $F_{prop}$ expressed by off-the-shelf, Vacco's MiPS and JPL hybrid thrusters (HT).}
\label{dragforce}
\end{figure}
From the literature, we found that propulsion systems for 1U commercial CubeSats can roughly express $F_{prop}=1mN$. For what concerns 6U CubeSats, we took as reference MarCO's platforms whose size well fits with the 6U model. MarCO satellites have been equipped with eight Vacco's thrusters but only four of them have been used for trajectory correction. Thus, from the data sheet of~\cite{Vacco} detailing the ``Micro Propulsion System" (MiPS) adopted by MarCOs, we fixed $F_{prop}=4\times 10mN$. However, thinking to Martian deep space missions with such a relevance, it would be useful to have more capacity for scientific payload. In this context, 12U CubeSat are the largest platforms from the considered ones. Without going much into the details, the authors in~\cite{HybridCS} presented a 12U CubeSat, weighing about $25kg$, with a single main hybrid rocket motor able to produce $F_{prop}=44.4N$, while occupying the $76\%$ of the total volume. This is more than $3$ orders of magnitude above the mentioned $F_{prop}$ for 6U CubeSat. 
Fig.~\ref{dragforce} shows, in a first instance, that such a system, with the current technologies, is not suitable for Earth. Indeed, by parameterizing Eq.~\ref{eq1} with $H_{Earth}=8.5km$ and $\rho_0=1.217\frac{kg}{m^3}$, we understand that the selected thrusters for the 1U, 6U and 12U CubeSat cannot provide enough force to correct trajectories and altitudes under, respectively, $[163.5,\, 143.0,\,86.5] km$. This is quite interesting but also expected due to the thinner Martian atmosphere with respect to Earth. This suggests that the atmosphere and the environment of the Red planet can be regarded as advantageous for the future \emph{in-situ} deployment of B5G networks. However, it is evident from Fig.~\ref{dragforce} that common commercial thrusters for 1U CubeSat, or the Vacco's MiPS for 6U CubeSat, cannot be used to guarantee Martian orbits with altitudes under $[134.5,\,108.0] km$, where our upper bound is $h_{max}\sim75km$. Instead, the JPL hybrid thruster can allow to decrease the minimum acceptable orbit well below $h_{max}\sim75km$, thus meeting the fundamental latency requirement of the splitting options 7.3, 7.2, 7.1 and 8. 12U hybrid CubeSat can support a minimum altitude $h_{min}\sim35km$. Clearly, it is not necessary to place constellations of CubeSats at such a low altitude, thinking also that it is almost impossible to maintain the thrusters always active to not consume the whole energy or propellant resources in a while. 
However, this gives us a consistent indication on the fact that in the next future, there could be Martian missions with extremely low altitudes. 

\begin{figure}[t]
\centering
\includegraphics[width=0.9\columnwidth]{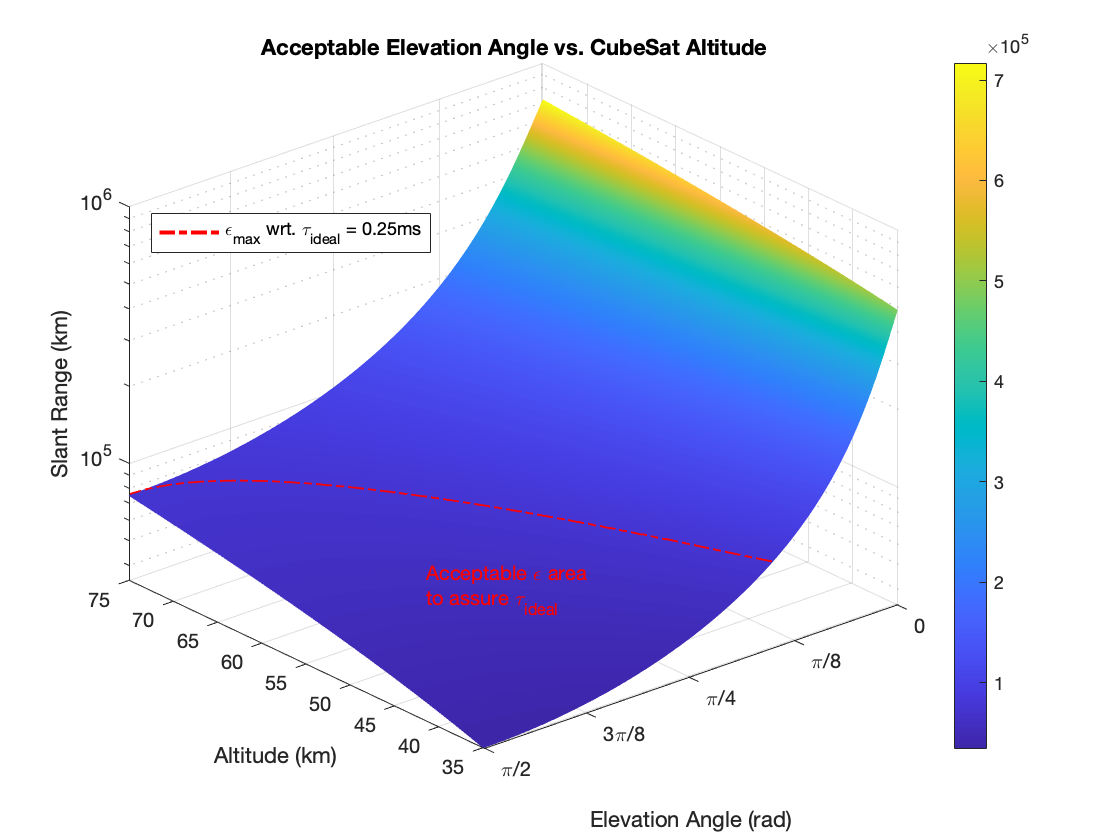}
\caption{3D plot representing the slant range vs. the altitude and the elevation angle. The red dotted line identifies the minimum elevation angle for each selected altitude.}
\label{elevangle}
\end{figure}

Our analysis proceeds by showing in Fig.~\ref{elevangle} the minimum acceptable elevation angle $\epsilon_{min}$ for altitudes ranging from the lower $h_{min}=35km$ and maximum $h_{max}=75km$ bound. As we can see by following the red dotted line, the slant range is fixed at $d_{max}=75km$ to respect $\tau_{ideal}=250\mu s$, i.e. the ideal latency case. The lower elevation angle $\epsilon$ is found when $h_{min}\sim35km$, while for $h_{max}=75km$ it would be possible a communication only at $\epsilon=\frac{\pi}{2}$, thus leading to a session time $t_{s}\sim0$. 
\begin{figure}[t]
\centering
\includegraphics[width=0.9\columnwidth]{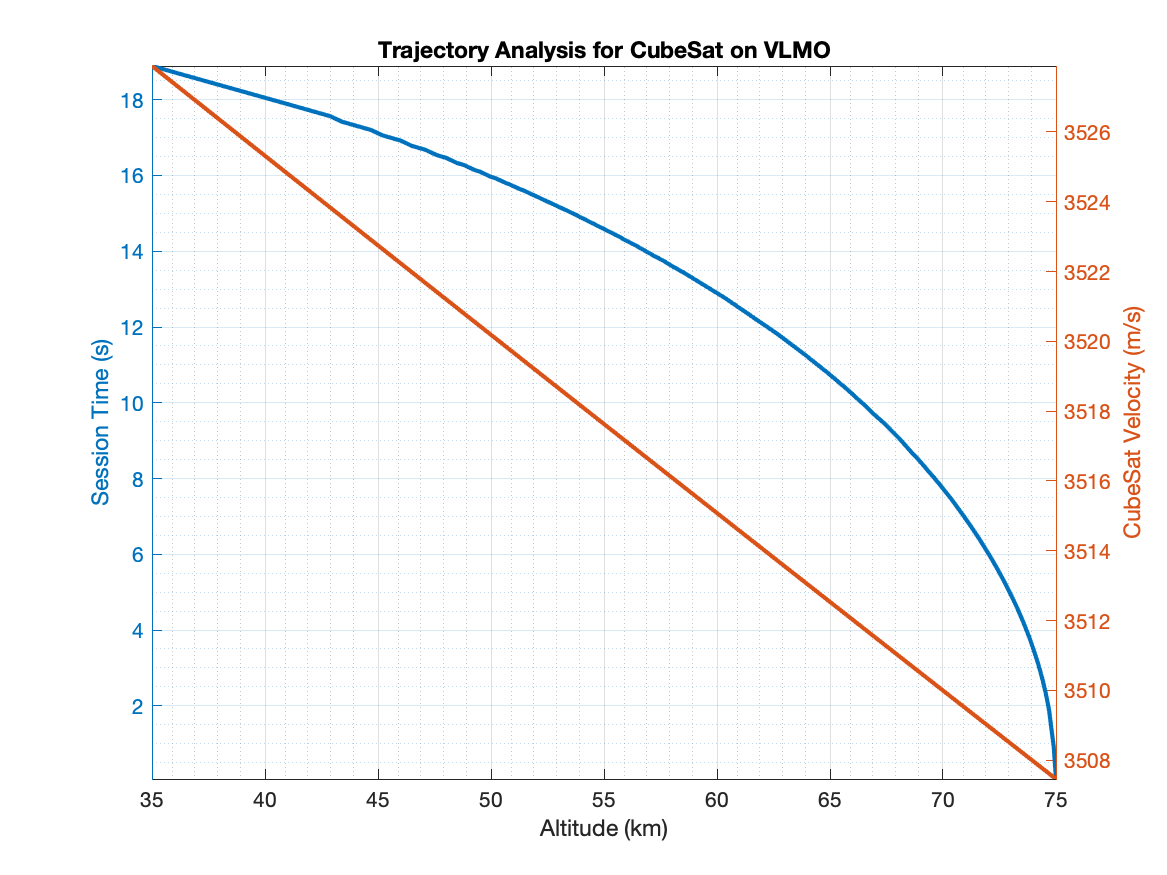}
\caption{For an altitude ranging from $h=[35,\,75]km$, the variation of CubeSat velocity and the available communication window from UAV to CubeSat.}
\label{commwindow}
\end{figure}
Now, Fig.~\ref{commwindow} depicts, for each altitude $h_{CS}$, the eventual session time between UAV and CubeSat, which is roughly moving at a speed $v_{CS}$ of about $\sim3.5\frac{km}{s}$. If we lower the altitude of CubeSat, we are able to sensibly increase the session time up to more or less $t_{s}=18s$ for really low altitudes, where we have to pay the price in terms of resources consumed to correct the trajectory of CubeSat.

\begin{figure}[t]
\centering
\includegraphics[width=0.9\columnwidth]{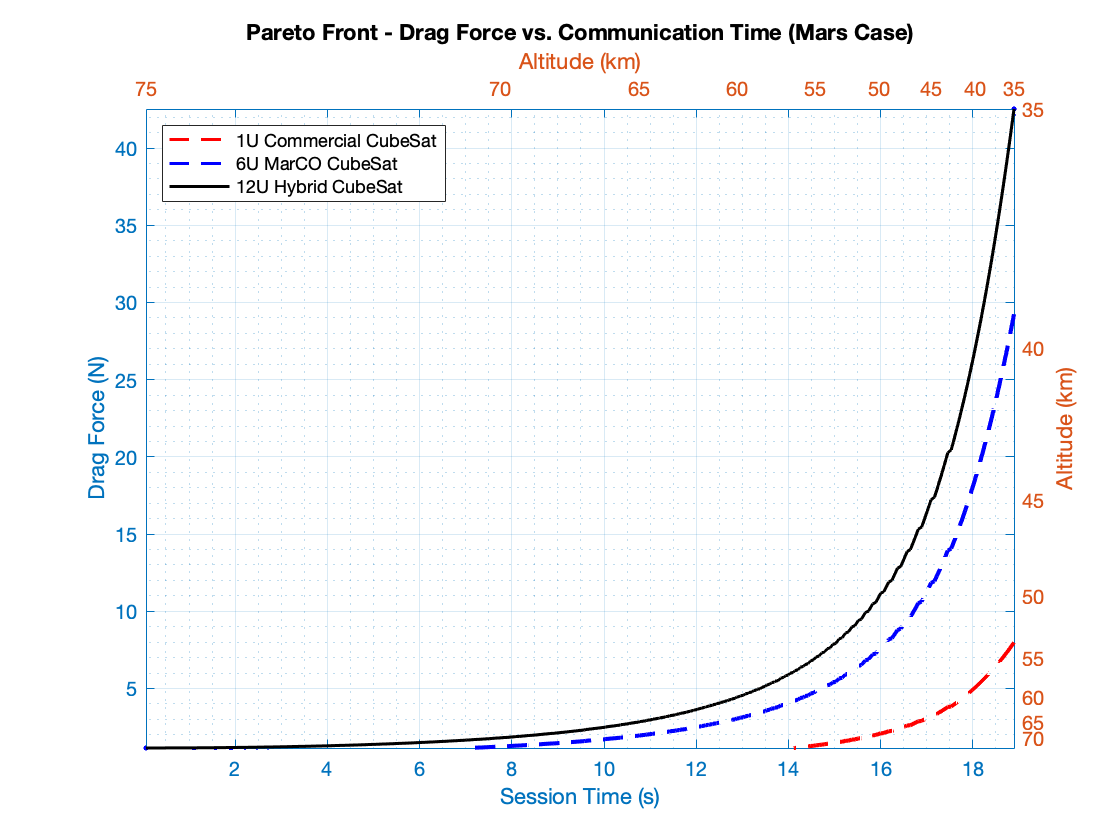}
\caption{Pareto front of the drag force $F_{drag}$ vs. session time $t_{s}$. The dotted lines refer to 1U and 6U CubeSats, while the black one is the 12U CubeSat.}
\label{pareto}
\end{figure}

To conclude, Fig.~\ref{pareto} directly correlates through a Pareto front the drag force and session time. As we higher the CubeSat altitude $h_{CS}$, we decrease the drag force $F_{drag}$ but also the possible session time $t_{s}$. Vice-versa, a lower $h_{CS}$ means higher $t_{s}$ but also higher $F_{drag}$ to be counteracted by $F_{prop}$ and the on-board resources. If we normalize the two terms, while giving them the same weights, and search for the altitude which minimizes the error, we obtain
\begin{equation}
\resizebox{1\hsize}{!}{$
    E=min\bigg(abs\Big(t_{s}(i+1)-t_{s}(i)\Big)+abs\Big(F_{drag}(i+1)-F_{drag}(i)\Big)\bigg)$}
    \label{eq8}
\end{equation} 
with $i$ the index running the vectors, 
the optimal altitude is $h_{opt}=67.1km$ with $F_{drag}=2.34N$ (the $5.2\%$ of the maximum $F_{prop}$ expressed by the JPL hybrid CubeSat) and $t_{s}=9.6s$. However, the selection of the best altitude should be done by unevenly weighing the objectives with respect to precise scientific requirements.

Now, let's consider the situation when we aim to avoid the thruster usage. In such a case, CubeSats should be placed into way higher LMOs with respect to those ones that we assumed to meet the ideal latency requirements for the functional split. Their orbital lifetime can be iteratively estimated, starting from a chosen altitude $h_{CS}$, an initial time $t=0$, Eq.~\ref{eq1} along with the relation between the orbital period $P$, and the semimajor axis (or radius $r$) of the circular orbit \cite{Decay}, i.e.:
\begin{equation}
    P=2\pi\sqrt{\frac{(h_{CS}+R_{Mars})^3}{G\cdot M_{Mars}}}
    \label{eq9}
\end{equation}
and the period decrease $dP$ caused by the atmospheric drag $F_{drag}$:
\begin{equation}
    \frac{dP}{dt} = -3\pi\rho(h_{CS})(h_{CS}+R_{Mars})\bigg(\frac{A_{CS}\cdot C_{D}}{m_{CS}}\bigg)     
    \label{eq10}
\end{equation}
where $m_{CS}$ is the CubeSat weight, which is shown in Tab.~\ref{tab:cs}. 
\begin{figure}[t]
\centering
\includegraphics[width=1.0\columnwidth]{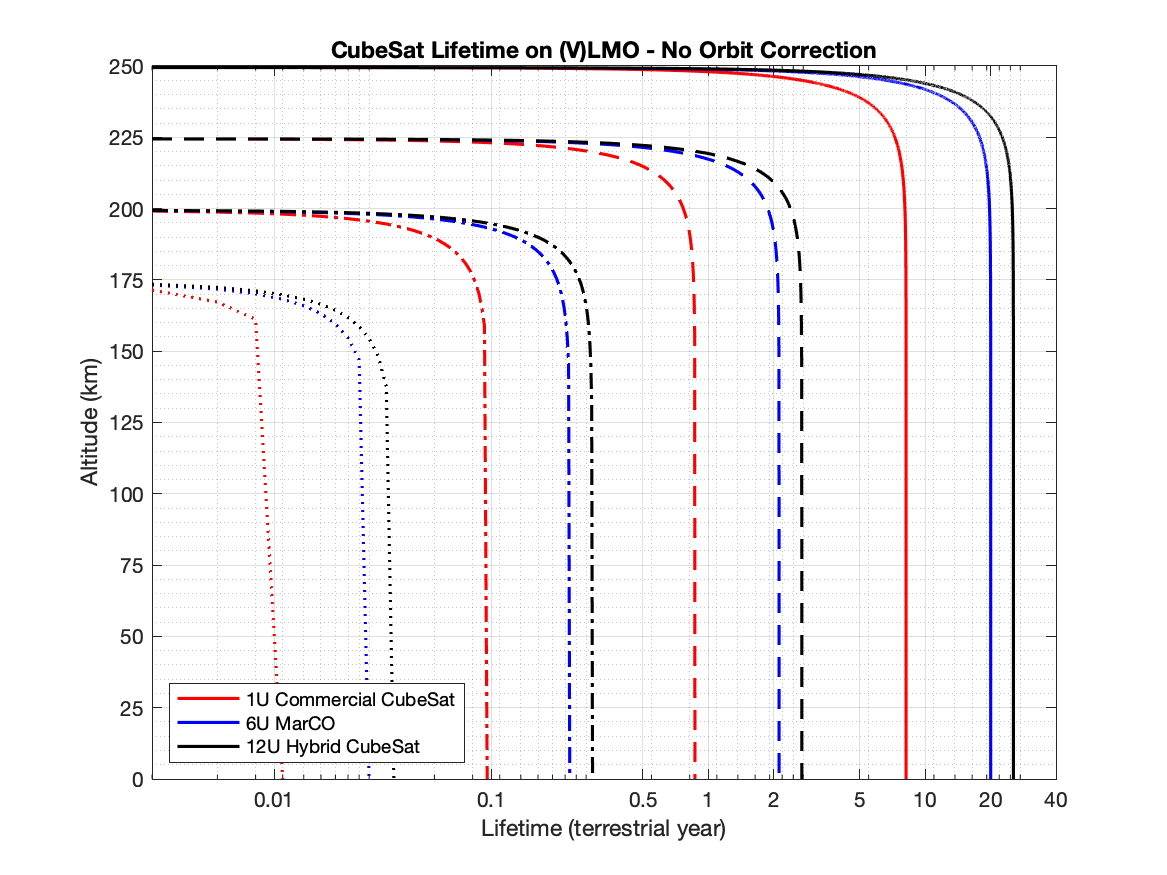}
\caption{\added{1U, 6U, 12U CubeSat lifetime without thrusters, thus no altitude correction over time.}}
\label{lifetime}
\end{figure} 
The orbital lifetime is the time iteration $t_{life}$ for which $P\left ( t_{life} \right )\rightarrow 0$. As shown in Fig.~\ref{lifetime}, the ratio between effective area and weight of CubeSats with the form factor 6U and 12U CubeSats helps to gain orbital lifetime with respect to 1U CubeSats. Indeed, as we lower the term $\frac{A_{CS}\cdot C_{D}}{m_{CS}}$, we lower the differential $\frac{dP}{dT}$, which is the main actor in the decrease of the orbital period. For $h_{CS}=75km$, without thrusters, a 12U CubeSats would fall within a couple of minutes, while for $h_{CS}=150km$ it would take just few more than a (terrestrial) day. As we move to higher orbits, such as $h_{CS}=175km$, a 1U, 6U an 12U CubeSat would definitely decrease their orbits, respectively, in 4, 10 and 13 days. On the other hand, an altitude of $h_{CS}=250km$ would lead to a 12U CubeSat orbital lifetime of about 25 years. However, the "Nanosats Database", probably the largest database of information regarding nanosatellites missions, evidences that 12U CubeSats have approximately an operational life of at least a couple of years \cite{NanoSat}. To guarantee 2 years of operational lifetime we should guarantee at least 2 years of orbital lifetime, which can be achieved by selecting $h_{CS}\sim225km$. Supposing $\epsilon=\frac{\pi}{2}$, thus $t_{s}\sim0$, the propagation delay will be around $\tau=0.75ms$. If we refer to the SCF latency requirements, splitting operations should be performed in the near-ideal case. This, for sure, would have some measurable impact on the overall performance of the 3D network. Further work will address this point by means of extensive network level emulations to assess many metrics, such as E2E delay, packet loss, throughput and achievable goodput. 

\section{Conclusion}
This paper presented the deployment of a Martian C-RAN using a 3D network architecture constituted by CubeSats and UAVs. We assessed the possibility of performing splitting opt.7-8 on non-dedicated hardware to divide the computational load of RU, DU and CU functions. Results quantitatively demonstrate the feasibility of the proposed architecture on Mars, at least in terms of latency requirements, provided that the lifetime and stability issues noticed in the early experiments of UAV and CubeSat launches in the Martian environment will be properly solved. Future work will concern with the end-to-end performance assessment of the considered Martian 3D networking system by using proper emulation tools. In such an analysis, the impact of relaxing latency and fronthaul rate requirements on network performance will be discussed in detail. Further work should also consider an in-depth analysis of the energy consumption of the various splitting options and the related impact on the design and future implementation of the proposed Martian 3D network architecture.

\section*{Acknowledgements}
The research activities presented in this paper fall within the field of interest of the IEEE AESS technical panel on Glue Technologies for Space Systems. This work is partly supported through a Faculty Fellowship awarded by DST NMICPS TiHAN at IIT Hyderabad and SERB Startup Research Grant (SRG-2021-001522).

\bibliographystyle{IEEEtran}
\bibliography{main}

\end{document}